\documentclass[final]{ws-ijmpb}
\usepackage{graphicx}
\usepackage{xspace}

\def\draftnote{} 

\begin{document}

\newcommand{\Htw}{H$_2$\xspace}
\newcommand{\Heth}{$^3$He\xspace}
\newcommand{\Hefo}{$^4$He\xspace}
\newcommand{\etal}{\textit{et~al}.\ }
\newcommand{\etalfin}{\textit{et~al}}
\newcommand{\cmref}[1]{cond-mat/#1}

\title{QUANTUM FLUIDS IN NANOPORES}
\author{NATHAN M. URBAN}
\address{Department of Physics, Pennsylvania State University\\University Park, Pennsylvania 16802-6300, USA\\nurban@phys.psu.edu}
\author{MILTON W. COLE}
\address{Department of Physics, Pennsylvania State University\\University Park, Pennsylvania 16802-6300, USA\\mwc@psu.edu}


\maketitle

\begin{abstract}
We describe calculations of the properties of quantum fluids inside
nanotubes of various sizes. Very small radius ($R$) pores confine
the gases to a line, so that a one-dimensional (1D) approximation
is applicable; the low temperature behavior of 1D \Hefo is discussed.
Somewhat larger pores permit the particles to move off axis, resulting
eventually in a transition to a cylindrical shell phase---a thin
film near the tube wall; we explored this behavior for \Htw. At even
larger $R\sim 1$~nm, both the shell phase and an axial phase are
present.  Results showing strong binding of cylindrical liquids
\Hefo and \Heth are discussed.
\end{abstract}

\section{Introduction}

The discovery of carbon nanotubes has provided a playground for
theoretical physics analogous to that ($\sim$1970) based on the
discovery of adsorption on flat, well-characterized surfaces. In
the former case, excitement arises from the tantalizing possibility
that one-dimensional (1D) physics can be tested by studying adsorbed
gases near nanotubes, just as studies of monolayer films provided
tests of 2D physics.

Many groups have explored the properties of quantum fluids on the
external surface of nanotube bundles and the interstitial regions
within the bundles, stimulated by both the intriguing geometry and
several experimental
results.\cite{ref1,ref2,ref3,ref4,ref5,ref6,ref7,ref8} Our
group has predicted several novel phase transitions for interstitial
quantum fluids, including a high temperature (liquid-vapor)
condensation and a BEC that exhibits 4D (!) thermodynamic
properties.\cite{ref2,ref3}  This paper summarizes instead diverse
results concerning quantum fluids \emph{inside} single nanotubes,
obtained with a variety of methods.  These studies are far from
complete, with significant theoretical questions yet to be answered.
The following section discusses the case of \Hefo in 1D, with
applications to small radius ($R$) pores.  Section 3 explores the
behavior of absorbed \Htw as $R$ increases, so that the 1D approximation
breaks down. Section 4 discusses the nature of films in large pores
($R\sim 1$~nm), where one encounters both a ``cylindrical shell''
phase of the film on the surface and a so-called ``axial'' phase,
which is very much like the 1D system in the small $R$ case.

Throughout this paper we omit the details of both the adsorption
potential and the techniques used in the calculations. Those can
be found in existing publications, as well as a thesis and longer
article currently being drafted.\cite{ref4} Our emphases are new
results, qualitative behavior, and significant open questions.

\section{Behavior in the 1D Limit}

The 1D \Hefo system is interesting for several reasons. One is that
the liquid is barely bound (by about 1.7~mK) and very rarefied (mean
spacing about 2.7~nm!); in fact, the venerable Lennard-Jones (LJ)
pair potential is too weakly attractive to produce this bound
state.\cite{ref5,ref6} A related fact is that the threshold
interaction strength for 1D binding of the liquid state coincides
with that of the 1D dimer.\cite{ref6} Recently, L.W.~Bruch and
C.~Carraro (private communications) have shown that the cohesive
energy of the 1D many-body system has the 1D dimer binding energy
as a lower limit, the two energies possibly coinciding.  There is
a closely related, intriguing aspect to the dimer problem.  Consider
the three-dimensional (3D) dimer problem, focusing on the $s$-wave
channel. The radial Schr\"odinger equation for that problem coincides
with the Schr\"odinger equation for the 1D dimer. A key difference
between $D=1$ and $D=3$ is the requirement that the wave function
$\psi(r)$ vanish at 3D separation $r=0$. However, this difference
is inconsequential for a 1D system involving hard-core interactions.
Hence, the 3D wavefunctions and spectra coincide, at least for the
$s$-channel, with those of the 1D problem.  Putting all of this
information together, it might be ``expected'' theoretically that
all of the three energies agree, with the common value 1.7~mK.

\begin{figure}
\centerline{\includegraphics[width=8.5cm,angle=-90]{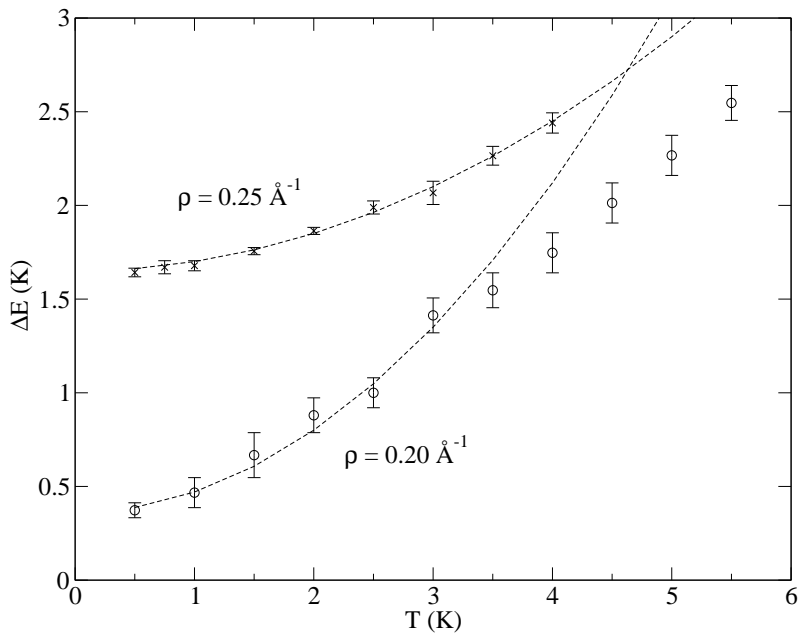}}
\caption{\label{fig:1}Energy per molecule of 1D \Hefo at $\rho=2.0$
and 2.5/nm.  Quadratic curves are based on the Landau-like model.
(From Ref.~\protect\refcite{ref4}.)}
\end{figure}

Recently, we have studied the thermal properties of 1D liquid \Hefo,
using the path integral method.\cite{ref17,ref19} If one were to anticipate
the behavior theoretically, one might treat the system with the
Landau model, based on elementary excitations above the ground
state. In the limit that the low-lying excitations are phonons,
with $T=0$ speed $s(\rho)$ at density $\rho$, this model predicts
that the energy per particle $\Delta E$ (relative to the ground
state energy) satisfies
\begin{eqnarray}
\Delta E(T) & = & F(k_B T)^2\,,\label{eq:1}\\
F & = & \frac{\pi/6}{\hbar s\rho}\,.\label{eq:2}
\end{eqnarray}
Preliminary results of the path integral calculations, in
Fig.~\ref{fig:1}, are consistent with this prediction at low $T$ and
high $\rho$; e.g., values of the coefficient $F$ in Eq.~\ref{eq:1}
fit to the data at $\rho=2.5$/nm (and higher) agree with the value
predicted by Eq.~\ref{eq:2}.  At $\rho=2$/nm and below, instead, the
values of $F$ begin to disagree and the departure from the quadratic
dependence occurs at lower $T$. The latter is not surprising because
the $T^2$ dependence of $\Delta E$ is expected only far below the
1D Debye temperature, $\Theta = \hbar s\rho/(k_B\pi)$.  Since
$\Theta\sim8$~K at $\rho=2$/nm and $\Theta\sim 30$~K at $\rho=2.5$/nm,
different behavior of $\Delta E$ is expected for the two densities
at 5~K.

Departure from simple model behavior at \emph{very} low $\rho$ is
also expected on general grounds, since the speed of sound becomes
imaginary below the spinodal density, $\rho\sim 0.245$/nm (compared
to an equilibrium density 0.36/nm). The phonon theory has no meaning
at such low $\rho$. The observed departure from Landau-like predictions
occurs at much higher density than that, however. This empirical
behavior remains to be understood.  At very low density, one might
be inclined to use a virial expansion.  {\v S}iber has evaluated
the first virial correction to the 1D \Hefo system. His results
show significant departure from the classical specific heat
($1/2$~Boltzmann per atom) at relatively high $T$, even for very
low density. This has nothing to do with exchange, which does not
contribute because of the hard core repulsion.\cite{ref20} One
expects higher order virial terms to contribute in addition, making
this system particularly interesting to explore experimentally.

\section{Spreading Away From the Axis}

The $T=0$ properties of 1D \Htw were first calculated by Boronat,
Gordillo, and Casulleras.\cite{ref7} Here, we report preliminary
results of finite $T$ behavior as the radius is increased from a
very small value to $R=0.7$~nm.

Assuming a Lennard-Jones interaction between the gas adsorbate
molecules and the inner pore surface, as adsorbate density is
increased the gas will begin to adsorb onto the pore as a cylindrical
shell film, at a distance from the surface on the order of the LJ
parameter $\sigma$.  However, we expect that if the radius $R$ of
the pore is less than this characteristic distance, the gas-surface
repulsion will heavily restrict the transverse motion of the gas,
and the two-dimensional (2D) shell will collapse into a 1D line of
molecules on the pore axis.

Geometrically, this axial compression at small pore radii exhibits
itself in the nanopore potential as a transition from a minimum
near $r\sim R-\sigma$ and a maximum at $r=0$ (i.e., an off-axis
potential well) when $R\gtrsim\sigma$, to a simple minimum at $r=0$
when $R\lesssim\sigma$ (an on-axis potential well).  For the case
of an infinitesimally thin tube, this transition is analytically
known to occur at $R\simeq 1.212\sigma$.\cite{ref14} We do not
know an analogous analytic result for the case studied here, a pore
within bulk material, but numerically the transition point is
similar.

For concreteness, we chose to study pores in MgO glass.  The \Htw-MgO
LJ interaction parameters are given by $\sigma = 2.014$~\AA,
$\epsilon=45.91$~K.\cite{ref15} The \Htw-\Htw interaction potential
was taken to be of the Silvera-Goldman form.\cite{ref16} The system
was studied numerically via the path integral Monte Carlo
algorithm.\cite{ref17,ref19} Temperatures between $T=0.5$~K and
3~K, radii between $R=2$ and 7~\AA, and densities of $\rho =
(2.55\times 10^{-3}/\textrm{\AA})/R^2$ and $\rho = (7.13\times
10^{-2}/\textrm{\AA})/R^2$.  (The densities correspond to choosing
a fixed number $N=1$ or $N=28$ of particles in a simulation cell
of radius $R$ and fixed length 125~\AA, with periodic longitudinal
boundary conditions.)

\begin{figure}
\centerline{\includegraphics[width=8.5cm,angle=-90]{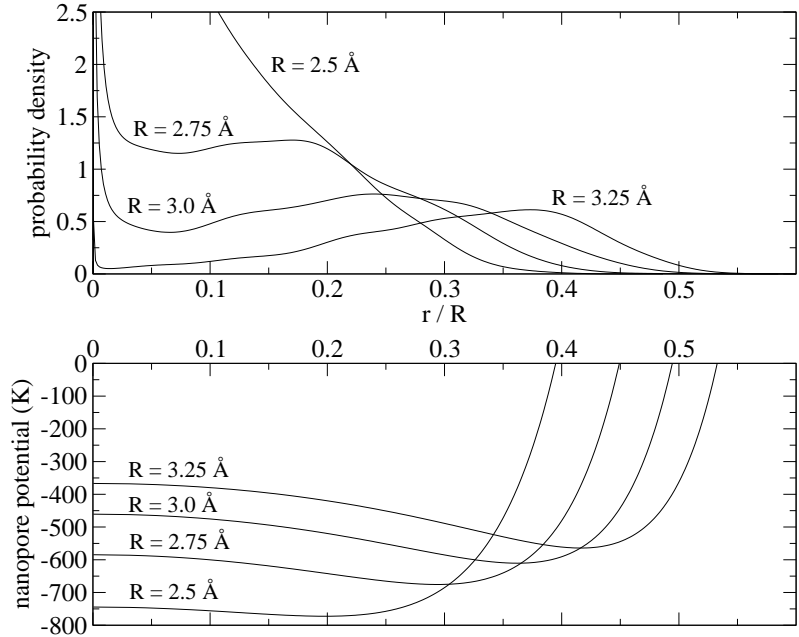}}\caption{\label{fig:2}
\Htw radial probability density distribution (at $T=0.5$~K, $\rho
= (2.55\times 10^{-3}/\textrm{\AA})/R^2$) and MgO pore potential,
for pore radii $R=2.5$, 2.75, 3.0, and 3.25~\AA, as functions of
dimensionless radius $r/R$.  (Radial densities near $r=0$ are
exaggerated due to finite size effects after normalizing the radial
distribution by $1/(2\pi r)$ to obtain the probability density.)}
\end{figure}

The results obtained were very similar for both densities and all
ranges of temperature and radii studied.  A representative plot of
the \Htw radial density distribution and the pore potential is given
in Fig.~\ref{fig:2} for radii between $R=2.5$ and 3.25~\AA.

The lowest pore size depicted, $R=2.5$~\AA, is just beyond the
transition point (2.45~\AA\@ for the tubular case) to an off-axis
minimum, where the \Htw should switch from a 1D axial phase to a
2D cylindrical shell.  This is reflected in the potential energy
curve (Fig.~\ref{fig:2}), which is very shallow for $R=2.5$~\AA.
Correspondingly, we see that at this radius the \Htw molecules are
still concentrated on the axis.  However, with a remarkably small
increase in pore radius, by 0.25--0.5~\AA, the distribution rapidly
shifts to peak off axis, signifying the onset of the shell phase.
It is interesting that the transition from axial to shell phases
coincides in $R$ with the shift of the classical potential minimum
from on- to off-axis, even in the presence of quantum effects and
interparticle interactions.  One expects larger effects of interactions
at higher density.


It should be noted that the radii depicted in Fig.~\ref{fig:2} are
unphysically small for MgO pores of realistic size.  We chose to
study them despite this problem in order to theoretically study an
axial to cylindrical shell transition in a pore geometry, which can
only occur for small pores which are highly confining.  In addition,
there are some systems of larger radii in which the first layer to
be adsorbed becomes a rigidly bound film, confining a fluid phase
to a very localized vicinity of the pore axis.\cite{ref18}  Such
possibilities help to justify the study of quasi-1D fluids.

\section{Large Pore Phenomena}

\begin{figure}
\centerline{\includegraphics[width=8.5cm]{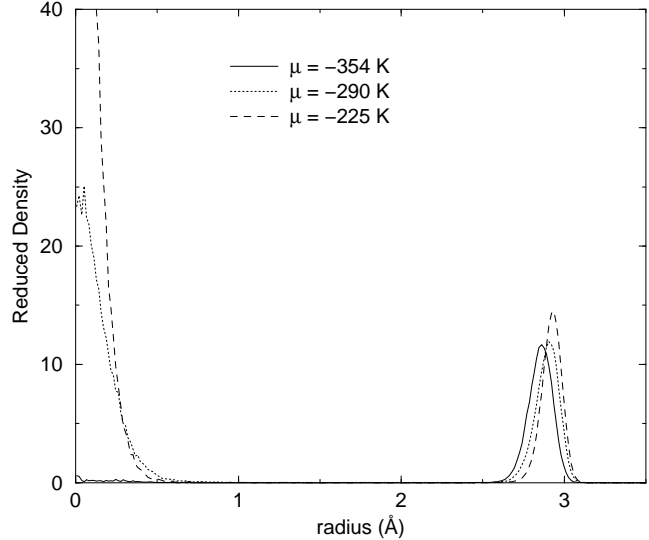}}
\caption{\label{fig:3} Pore-filling transition of \Htw in a tube
of radius 0.6~nm, from Gatica \etalfin.\protect\cite{ref10} Results
are densities as a function of $r$ at 10~K and indicated chemical
potentials.}
\end{figure}

Relatively few simulation studies have been carried out for quantum
fluids in ``typical'' size nanotubes, $R\sim 0.6$ to 1~nm. Path
integral calculations of Gatica \etalfin.\cite{ref10} reported the
behavior of \Htw over a limited range of $R$ at $T=10$~K. One of
the more interesting phenomena is the pore-filling transition, shown
for \Htw in Fig.~\ref{fig:3}.\cite{ref10} At low $\mu$, all of the
molecules are localized within a thin layer, at $r=0.3$~nm, located
near the distance of closest approach to the nanotube.  Above a
threshold value of $\mu$, the axial phase appears and grows rapidly
with increasing $\mu$. This axial phase can be thought of as an
independent 1D phase. A close analogy is the behavior of the second
layer film of He or \Htw on the surface of graphite, often treated
by assuming that the only role of the first layer is to provide a
holding potential.\cite{ref11}

\begin{figure}
\centerline{\includegraphics[width=8.5cm,angle=-90]{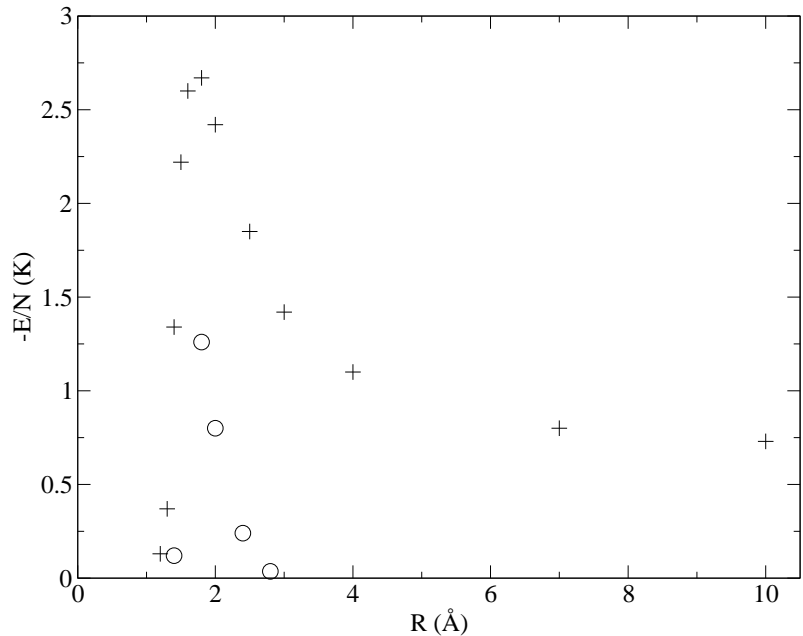}}
\caption{\label{fig:4} Cohesive energy per atom of cylindrical
liquids, as a function of $R$. Pluses are \Hefo data (from Kostov
\etalfin.\protect\cite{ref12}) and circles are \Heth data (from
C.~Carraro, unpublished).}
\end{figure}

Analytical and numerical problems associated with matter in cylindrical
geometry are often computationally demanding, motivating the use
of simplifying models that (we hope) capture the essential physics.
Recently, we have explored such a description of the shell phase;
the model assumes that all particles are constrained to lie on a
cylindrical surface, $r=R$. One might expect that by varying $R$
between $R=0$ and $R=\infty$, one interpolates smoothly between 1D
and 2D behaviors.  This is na\"ive; instead, an intriguing ``anomaly''
arises: a significant enhancement of the binding occurs when the
diameter of the cylinder, $d=2R$, is comparable to the equilibrium
separation $r_\mathrm{min}$ in the pair potential. The condition
$d\sim r_\mathrm{min}$ corresponds (for LJ interactions) to
$\sigma/R\sim 1.7$.  Indeed, this argument does explain the $R$
value corresponding to the maximum cohesive energy (seen in
Fig.~\ref{fig:4}) of the ``cylindrical liquids'' \Hefo and \Heth.
The \Heth case is perhaps the most dramatic, because its liquid
state does not exist in either 1D or 2D, while the cylindrical
liquid \Heth is found to have cohesive energy as high as 1.26~K for
$R=0.18$~nm.

These are variational results, obtained with Jastrow and Slater-Jastrow
wave functions for \Hefo and \Heth, respectively. Qualitatively
similar, enhanced binding behavior was found for related problems
involving similar binding problems on a cylindrical surface: He or
\Htw dimers, a crystalline lattice confined to a cylinder, and the
virial coefficient of a classical fluid.\cite{ref12} The origin
of this general behavior is that two interacting particles can
maximally exploit a divergent ``specific area'' when the interatomic
separation is favorable.  This occurs when the particles are on
opposite sides of the cylinder, with separation
$|\mathbf{r}_2-\mathbf{r}_1|=d=r_\mathrm{min}$.  The specific area
is defined as the cylindrical area residing within a separation
interval $[r, r+dr]$, divided by $dr$. For completeness, we note
that analogous results for the dimer binding have been found by
Aichinger~\etalfin., using both the simple model of confinement on
a cylinder and more realistic study of dimers inside a
nanotube.\cite{ref13} The optimal binding value found for $R$ is
very different in the two cases.

\section*{Acknowledgments}

We are grateful to our collaborators (Massimo Boninsegni, Louis
Bruch, Mercedes Calbi, Carlo Carraro, Silvina Gatica, and Milen
Kostov) for many contributions to this project and to NSF for its
support of this research.

\end{document}